\documentclass{article}

\usepackage[space]{grffile} \usepackage{latexsym}
\usepackage{textcomp} \usepackage{longtable}
\usepackage{multirow,booktabs} \usepackage{amsfonts,amsmath,amssymb}
\usepackage{cite}

\usepackage{url} \usepackage{hyperref}
\hypersetup{colorlinks=false,pdfborder={0 0 0}}

\usepackage[affil-it]{authblk}

\usepackage{graphicx,color,epsfig} \usepackage{booktabs}
\usepackage{colortbl} \newcolumntype{C}{c<{\kern\tabcolsep}@{}}
\usepackage[table]{xcolor} \usepackage{stfloats}
\usepackage[utf8]{inputenc}

\definecolor{Gray}{gray}{0.9} \definecolor{LGray}{gray}{0.7}
\definecolor{Blue}{RGB}{175,238,238}
\definecolor{LBlue}{RGB}{224,255,255}

\graphicspath{{./Figs/}}

\begin{document}
\title{Performance of attack strategies on modular networks}

\author{Bruno Requião da Cunha} \affil{Polícia Federal,
  Brazil\\ Instituto de Física, Universidade Federal do Rio Grande do
  Sul, Porto Alegre, RS, Brazil}

\author{Sebastián Gonçalves} \affil{Instituto de Física, Universidade
  Federal do Rio Grande do Sul, Porto Alegre, RS, Brazil}

\date{\today}


\begin{abstract}
Vulnerabilities of complex networks have became a trend topic in
complex systems recently due to its real world applications. Most real
networks tend to be very fragile to high betweenness adaptive
attacks. However, recent contributions have shown the importance of
interconnected nodes in the integrity of networks and module-based
attacks have appeared promising when compared to traditional malicious
non-adaptive attacks. In the present work we deeply explore the
trade-off associated with attack procedures, introducing a generalized
robustness measure and presenting an attack performance index that
takes into account both robustness of the network against the attack
and the run-time needed to obtained the list of targeted nodes for the
attack.  Besides, we introduce the concept of deactivation point aimed
to mark the point at which the network stops to function properly.  We
then show empirically that non-adaptive module-based attacks perform
better than high degree and betweenness adaptive attacks in networks
with well defined community structures and consequent high modularity.
\end{abstract}

\maketitle

\section{Introduction}\label{intro}

Structural vulnerabilities of real systems have attracted much
attention from the network science community
recently~\cite{newmanbook,havlincomplenet} both from the attack point
of view (when we are interested in disabling or fragmenting a network
with as little effort as possible)~\cite{Holme:2004aa} and from the
security point of view (when we wish to created safer networks or to
defend them against targeted or malicious attacks)~\cite{Wang:2007}.
For instance, the operation of internet routers~\cite{Cohen2001}, the
delivery of drugs in biological systems, the propagation of an
epidemic disease~\cite{Hebert-Dufresne:2013aa}, the security of a
power grid~\cite{Pagani20132688,Habib:15}, or even the operativeness
of organized crime or terrorist cells~\cite{Duijn2014} are all
examples of networking systems in which we are much interested either
in devising efficient attack strategies to rapidly atomize the
network~\cite{Raab:2003aa} or in adopting defensive actions to prevent
the system from collapsing~\cite{albert2004structural}.

Networks might be structurally affected either by random removal of
nodes (failures) or by targeted or malicious
attacks~\cite{Crucitti2003,Albert2000,Crucitti2004,Jeong2001}.
Targeted attacks are usually aimed to disrupt the system by removing a
small fraction of nodes or edges.  In this sense, traditional attack
methods usually focus on the sorting of nodes according to their
importance in the network architecture, \emph{i.e.} according to some
centrality index ---betweenness and degree-based attacks usually
present better results~\cite{Iyer2013}.

Basically there are two approaches for network fragmentation:
non-adaptive (or simultaneous) attacks and adaptive (or sequential)
attacks. In the first approach, the list of attacked nodes or edges is
produced only once, before the removal procedure
starts~\cite{Iyer2013}. In the
second~\cite{holme2002attack,Buldyrev2010} approach, the list of
targets is updated after each deletion by recalculation of the
centrality index used to sort nodes or edges.  Consequently, high
adaptive attacks demand more processing time, but on the other hand
the method usually produces more damage per removal when compared to
the non-adaptive approach.  The reason is more or less obvious: if the
list of attack is measured only once, the method cannot account for
the changes in the centrality order due to the removal of elements.
So, in the worst case the high adaptive version of a procedure is as
good as the non-adaptive approach, however it is generally better.

Nonetheless, real networks tend to organize into modular structures or
communities ---clusters densely connected internally but sparsely
connected among them~\cite{newmanmod}--- and recently it was shown
that the nodes bridging communities are even more crucial in keeping
networks from falling apart than highly connected
vertices~\cite{shai2014resilience,havlinshairesilience}.  Even more
recently, it was shown empirically that module-based attacks (MBA)
targeting interconnected nodes (communities bridge) can damage real
networks with more efficiency than other non-adaptive targeted
attacks~\cite{fast}.

Hereupon, although there has been much work recently on network
robustness, the balance between a given network damage and the
computational cost (run time and/or hardware capacity) necessary to
reach that desired level of fragmentation is an issue not yet properly
addressed by previous contributions.

Usually the only aspect taken into account is the efficiency in terms
of the ratio between the damage produced and the fraction of removed
nodes. In this contribution, apart from that feature, we consider the
trade-off between the network robustness and the computational cost of
a given network attack strategy. More precisely, we focus on the
cost/benefit relation of adaptive and non-adaptive attacks to modular
networks in order to rightly chose the most appropriate strategy to
atomize large complex networks.

\section{Module-based attacks}\label{mba}
The MBA procedure mentioned above is a non-adaptive attack which
consists of targeting interconnected nodes ordered by betweenness
centrality. Besides, once a node from a edge between two connected
modules is deleted, there is no need to erase its counterpart unless
it still connects to other communities ---the vertices belonging to
this set are called independent interconnected nodes. Furthermore, the
attack is aimed at the largest remaining connected component of the
network after each step. As detailed in \cite{fast} in module-based
attacks for real networks, the communities are usually extracted by an
heuristic community detection algorithm due to the size of such
systems and consequent computational requirements. Besides, the
effectiveness of the MBA procedure is closely related to the
modularity of the network.

The modularity of an unweighted network is usually defined as the
density of links inside communities as compared to links between
communities, as follows~\cite{newmanbook}:
\begin{equation}\label{modularity}
Q = \frac{1}{2m}\sum_{i,j} \left[A_{ij} -
  \frac{k_ik_j}{2m}\right]\delta(c_i,c_j)\,
\end{equation}
where $A_{ij}$ is the adjacency matrix (taking the value $1$ when
there is a link between nodes $i$ and $j$, $0$ otherwise), $k_{i}$ is
the vertex degree of node $i$ and $c_{i}$ represents the community to
which this node belongs. The $\delta$-function $\delta(u, v)$ is $1$
if $u = v$, $0$ otherwise and $m$ is the total number of edges.

In this sense, many methods for community detection have been devised
over the last few years. Therefore, the issue of testing the accuracy
of community extraction algorithms is crucial in studying modular
networks. In order to test the performance of algorithms for community
extraction (or identification), several benchmark for computer
generated networks, with well-defined community, structures have been
proposed. One of the first introduced benchmarks for testing
community detection is a class of artificial undirected networks
proposed by Girvan and Newman (GN)~\cite{gnbench}. It consists of
networks with nodes having approximately the same degree, as in a
random graph, but with nodes preferentially connected to nodes of
their group.
%
However, real networks have heterogeneous distributions of node degree
and community size accounting for several remarkable features of real
networks, such as resilience to random failures/attacks and the
absence of a threshold for percolation and epidemic spreading. In this
sense Lancichinetti, Fortunato, and Radicchi have proposed an
undirected network benchmark (LFR)~\cite{lancichinetti2008benchmark}
which assumes that both degree and community size distributions are
power laws.

In these benchmarks, the modularity is controlled by the mixing
parameter $\mu$, which is the ratio between the number of edges
linking a vertex to other communities and its degree. In other words,
each node shares a fraction $\mu$ of its edges with nodes of the entire network and 
a fraction $1-\mu$ of its edges with nodes
of its own community.  Thence, small values of $\mu$ indicate well separated
module, whereas higher values means communities are ill defined and
possibly overlapped.  It should also be noted from now on that the
average module size usually depends on the accuracy and the detection
threshold of the community extraction algorithm. However, as pointed
out in ~\cite{lancichinetti2009community} the \emph{Louvain} algorithm
by Blondel \emph{et al.}~\cite{blondel2008fast} is known to perform
very precisely in both GN and LFR benchmarks. Therefore, in
simulations hereafter we use only the \emph{Louvain} method to detect
communities.

\subsection{Deactivation point and the generalized robustness}
Even though there has been many works recently concerning the
robustness of complex networks, there is not an unique definition of
it ~\cite{Iyer2013,Crucitti2003,Albert2000,Crucitti2004}. Robustness
might be defined as the ability of a system to keep a set of defined
features or services working when subject to perturbations or
attacks~\cite{jen2005robust}.  This response to perturbations is
tightly coupled to the goal of the real system subjacent to the graph
representation in such a way that the robustness is service
dependent~\cite{van2010framework}.  Therefore, the robustness of
complex abstract graphs, in which the system dynamics is not taken
into account, should be addressed by either topological phase
transition points or, in the absence of such behavior, by general
functions representing the overall response of the network topology to
a desired strategy of attack or fragmentation~\cite{Schieber2016359,
cohen2000,Callaway2000,pu2015vulnerability,Pu2012,Boginski2009}.
In this context, robustness is typically considered in a percolation
framework and quantified by the critical fraction $\rho_{c}$ of nodes
that once removed by degree-based attacks leads to complete
atomization of the network. At that point, one may safely say that the
network stops functioning as whole because there is no giant component
connecting the system according to the Molloy-Reed criterion.  In this
framework, in order to quantify the effect of the attacks on the
networks~\cite{barthelemy2011spatial}, it is usually defined the order
parameter $\sigma (\rho )= \frac{N_{\mathcal{L}}}{N}$ as the relative
size of the remainder network size relative to the original network
size as a function of the fraction of nodes deleted.


However, modular networks usually present a transition point that
marks a phase where all communities are detached from the main
graph. At this point, the largest cluster size would be equal to the
original network's maximum module size $N_{max}$ and the critical
fraction of nodes that should be deleted in order to stop the normal
operation of the network is given by what we call the deactivation
point $P_{d}=(\sigma_{d},\rho_{d})$, where
\begin{equation}\label{deactivationpoint}
\sigma_{d}=\frac{N_{max}}{N},
\end{equation}
is the size of the largest module as compared to the original network
size and $\rho_{d}$ is the fraction of independent interconnected
nodes.  This point depends on the modular structure of each network
and after this phase is reached, communication would not be possible
among large distant pieces of the network. In other words, this point
marks a phase at which the network effectively stops functioning as a
whole much earlier than the percolation threshold under hub attack and
before the network atomizes completely.

On the other hand, in order to compare the deactivation threshold
($\rho_{d}$) of networks with different topologies, the size of the
largest community ($\sigma_{d}$) would have to be similar for most
network types. This is not the case as easily observed in
Eq.~\ref{deactivationpoint}. Besides, the attack set should be the
same if we are interested in comparing different networks and multiple
attack methods. Again, this is not case because module-based attacks
usually generate attack lists much smaller than the network
size. Therefore, we need a more general robustness approach to account
for situations where the network loses its essential modular
functionality before it collapses completely after a given number of
nodes, much smaller than the original size of the network, is
removed. In this sense, a more general approach considers the relative
size of the largest remaining connected component of the network
during the attack procedure.

\begin{figure}
\begin{center}
\includegraphics[width=0.7\columnwidth]{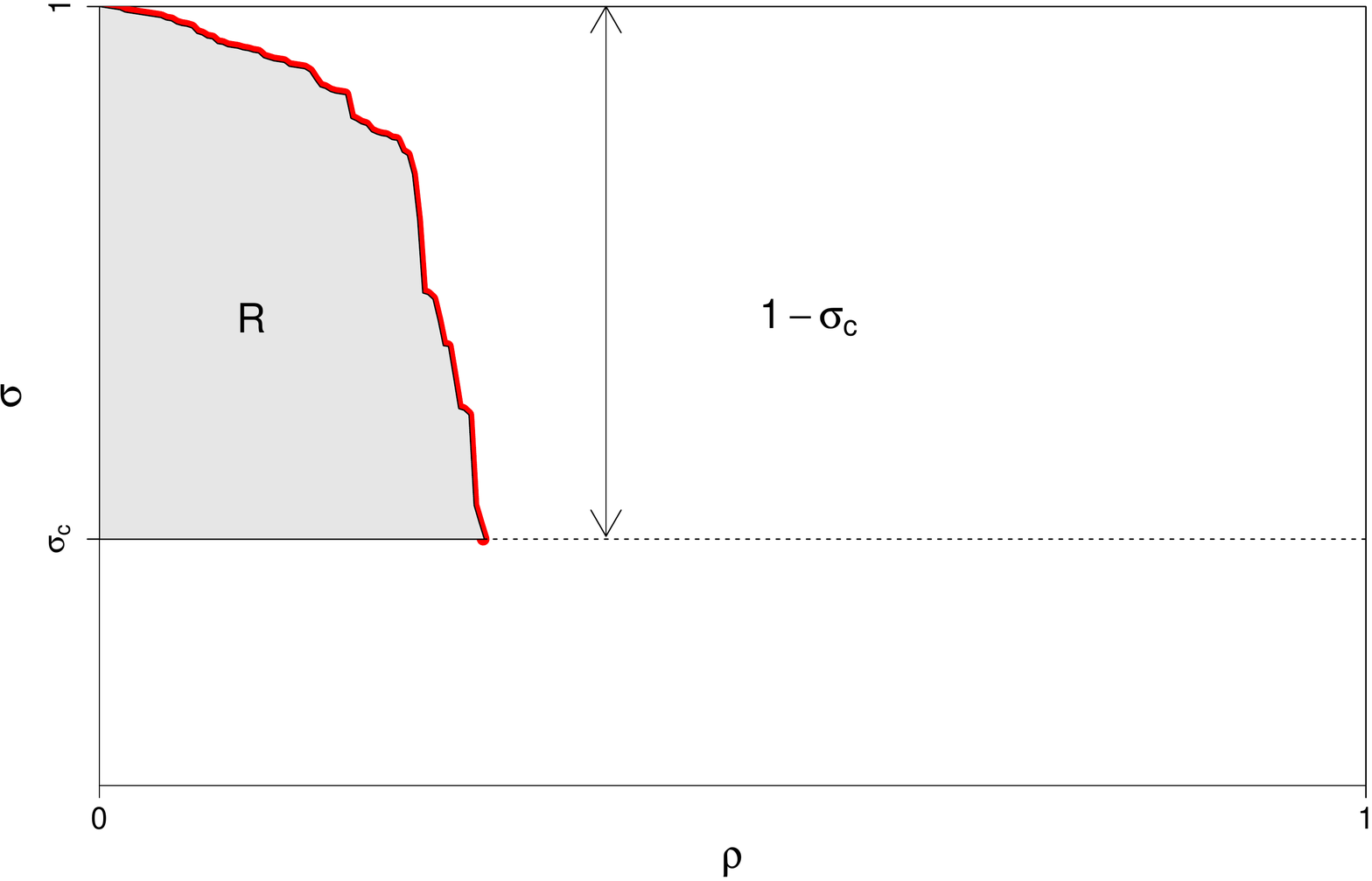}
\caption{The figure depicts the geometric representation of the
  robustness as the ratio between the area underneath the red curve
  and the total area of possible attack given by area of the rectangle
  with sides lengths $1-\sigma_{c}$ and $1$ as defined in
  Eq.~\ref{rob}.\label{robustness_def}}
\end{center}
\end{figure}

Soares \emph{et al.} have proposed an unique measure to quantify the
robustness of a system facing malicious attacks~\cite{Schneider2011},
that resumes to the area of $\sigma(\rho)$ curve. Here, we generalize
this concept for targeted attacks that end before all nodes are
removed, such as the module-based attack. In this sense, the
robustness of a network to a particular attack strategy is given by 
(see Fig.~\ref{robustness_def} for details):
\begin{equation}\label{rob}
R=\frac{1}{N(1-\sigma_{min})}\sum_{\rho=0}^{\rho_{max}}\sigma(\rho)
\end{equation}
where $\sigma$ is the size of the largest connected component relative
to the original size of the network, $N$ is the number of nodes in the
network, $\rho$ is the fraction of nodes removed, $\rho_{max}$ is the
point at which the attack ends and $\sigma_{min}$ is the value of the
relative size of the largest connected component at $\rho_{max}$. This
quantity measures the area underneath the $\sigma$ curve relative to
the maximum area of attack, \emph{i.e.} the area of the rectangle
delimited by the points ${(0,\sigma_{min}), (0,1), (1,\sigma_{min}),
  (1,1)}$. In the MBA case, the point $(\rho_{max},\sigma_{min})$ is
the deactivation point $(\rho_{d},\sigma_{d})$.

\subsection{Run-time and attack performance}
Usually, the most used and efficient known methods of adaptive network
attack are the high degree adaptive attack (HDA) and high betweenness
adaptive attack (HBA).  Therefore, from now on we compare the MBA
procedure to these two methods.  In order to study the connection
between robustness and computation time we must calculate the time
needed to perform each attack: HBA, HDA and MBA.  Computing the
betweenness centrality of all nodes in an unweighted network usually
takes $\mathcal{O}(NE)$ time~\cite{Brandes:2001aa} while the degree
complexity has a linear dependence
$\mathcal{O}(N+E)$~\cite{brandesbook}. On the other hand, even though
the exact computational complexity of the \emph{Louvain} method is not
known, it seems to run in time $\mathcal{O}(N\log
N)$~\cite{blondel2008fast}.  Therefore, the computational run-time of
MBA,HDA and HBA attacks are in increasing order of complexity
$\mathcal{O}(NE+N\log N)$,
$\mathcal{O}(\sum{(\mathcal{N}+\mathcal{E})})$ and
$\mathcal{O}(\sum{\mathcal{N}\mathcal{E}})$. As expected, MBA (even
for dense graphs) is less expensive computationally than sequential
methods, followed by HDA and HBA for a given network size.

In this sense, we may define the performance of an attack to a given
network by:
\begin{equation}\label{eq1}
\mathcal{P}=\frac{1}{t\times R}
\end{equation}
where $t$ is the time taken to complete the procedure in seconds and
$R$ is the robustness. In other words, $\mathcal{P}$ measures the
trade-off between the network robustness to a given attack and the
time taken to complete the attack-- a fast computing algorithm that
efficiently fragments a network should result in high values of the
attack performance $\mathcal{P}$, while very efficient attack methods
that are in turn very slow should have attenuated performance values.

\section{Results}
We now generate multiple benchmark networks with varying size and
modular structure according to Table~\ref{tab_data}. For both classes
of undirected benchmarks (GN and LFR) the mixing parameter varies as
$0.05<\mu<0.3$ and the modularity as $0.62<Q<0.93$.  The next step is
to analyze both the performance ($\mathcal{P}$) and the network
robustness ($R$) according to each of the following fragmentation
prescriptions: high degree adaptive attack (HDA), high betweenness
adaptive attack (HBA) and module-based attack (MBA).

\begin{table}[h]
\centering
\begin{tabular}{lp{0.4in}p{0.4in}p{0.4in}p{0.4in}}
\textbf{Net} & \textbf{N} & \textbf{E} & $\mu$ & \textbf{Q}\\\hline \hline
 	GN1  & 1000 & 2500 & 0.05 & 0.93\\ 
  	GN2  & 1000 & 2500 & 0.10 & 0.87\\ 
 	GN3  & 1000 & 2500 & 0.15 & 0.81\\ 
  	GN4  & 1000 & 2500 & 0.20 & 0.76\\ 
  	GN5  & 1000 & 2500 & 0.30 & 0.62\\
	\hline
 	LFR1 & 1000 & 2286 & 0.05 & 0.91\\
  	LFR2 & 1000 & 2385 & 0.10 & 0.86\\ 
	LFR3 & 1000 & 2314 & 0.15 & 0.83\\ 
 	LFR4 & 1000 & 2392 & 0.20 & 0.76\\ 
 	LFR5 & 1000 & 2292 & 0.30 & 0.68\\
	\hline
 	EUPG  & 1494 & 2322 & -      & 0.89\\\hline\hline
\end{tabular}
\caption{In this table we present the topological data for the five artificial networks 
of the Girvan-Newman class (GN1 - GN5), the five artificial networks of  Lancichinetti-Fortunato-Radicchi class (LFR1 - LFR5) 
and the European power grid system (EUPG). The data consists of the network type, the 
number of vertices, the number of edges, the mixing parameter and the modularity.}
\label{tab_data}
\end{table}

\begin{figure}
\begin{center}
  \includegraphics[width=1\columnwidth]{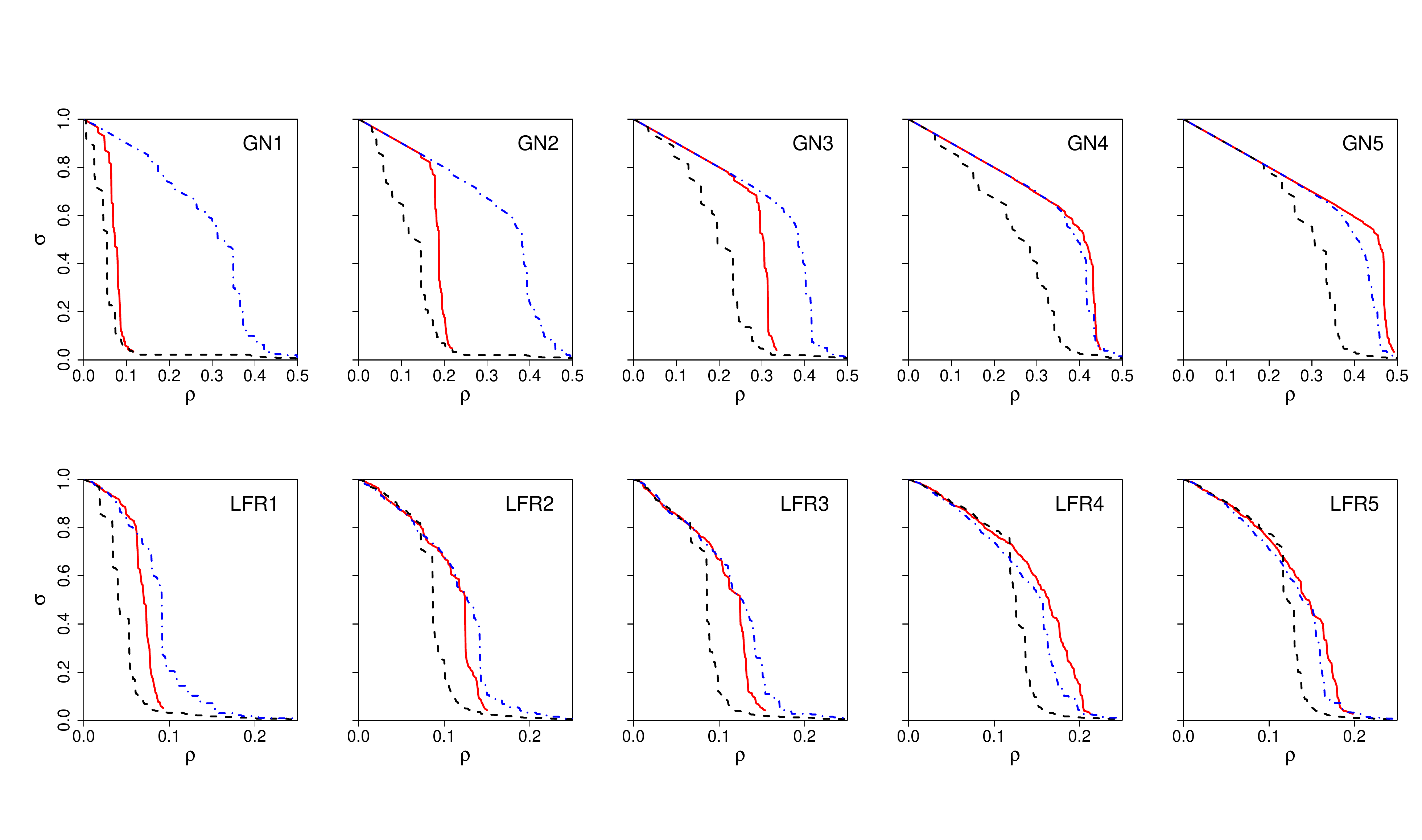}
  \caption{The fragmentation process presented by the relative size of the biggest component $\sigma$ 
  as a function of the fraction of removed nodes $\rho$ following MBA, HBA
  and HDA attacks on the benchmark networks, Girvan-Newman (GN) and
  Lancichinetti-Fortunato-Raddichi (LFR) defined in Table~\ref{tab_data}. 
  Modularity decreases from left to right. Solid red lines are MBA attacks, 
  dashed black lines are HBA attacks and dashed-dotted lines are HDA attacks.
\label{benchmarks}}
\end{center}
\end{figure}

As can be seen in Fig.~\ref{benchmarks}, networks with high modularity
tend to be less robust to MBA attacks as expected. However, a new 
important feature of the MBA becomes clearer. 

\begin{figure}
\begin{center}
\includegraphics[width=0.7\columnwidth]{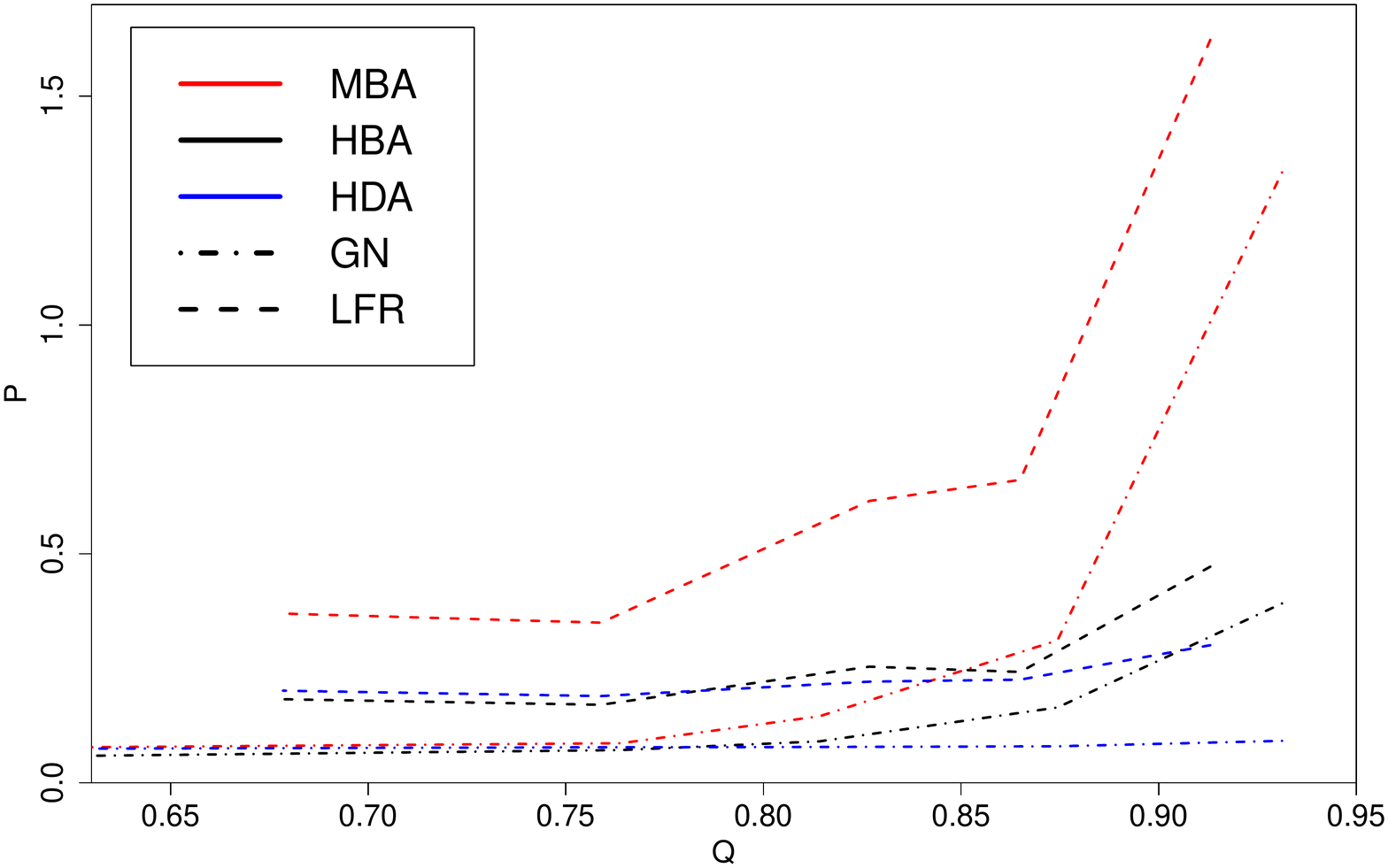}
\caption{The figure shows the performance $\mathcal{P}$ defined 
Eq.~\ref{eq1} as a function of the modularity $Q$ for the ten 
networks detailed in Table~\ref{tab_data}. Red lines depict MBA 
attacks, blue lines HBA attacks and black lines HDA attacks. 
The benchmarks are divided with the dotted lines representing 
Girvan-Newman networks and the dashed lines the 
Lancichinetti-Fortunato-Radicchi benchmarks.\label{performance}}
\end{center}
\end{figure}

As the modularity of the
networks is increased simultaneous MBA strongly outperforms adaptive
degree attacks and more, for highly modular networks the robustness to
MBA and high betweenness adaptive attack are very similar, with the
MBA approach being much faster computationally than the sequential
approach. This feature is easily seen by the network performance
defined by Eq.~\ref{eq1}.  In Fig.~\ref{performance} we plot
$\mathcal{P}$ with time measured as CPU time in seconds. Results show
that the MBA approach has a better trade-off between efficiency and
computational cost than adaptive methods. For the chosen LFR
benchmarks the MBA performance always outperforms HBA and HDA
performances, while for the GN benchmarks there is a critical
modularity of approximately $0.7$ from which $\mathcal{P}(MBA)$ is
higher than in the adaptive approaches.

As a final case study, we address a real network representing the
European power grid (EUPG)~\cite{Shavitt:2010aa}. This system consists
of 1494 nodes and 2322 edges and a node represents a generator, a
transformer, or a substation, while edges represent a power supply
line. The European power grid network is highly modular with a
modularity extracted by the \emph{Louvain} method of $0.89$. As shown
in Fig.~\ref{eupg} and its inset, the performance of HDA,HBA and MBA
attacks in the European power grid network are respectively
$\mathcal{P}_{EUPG}(HDA)=1.59$, $\mathcal{P}_{EUPG}(HBA)=2.73$ and
$\mathcal{P}_{EUPG}(MBA)=15.65$. As expected by the analyzes on the
benchmarks, in the power grid system the performance of the MBA
prescription is much higher than performance HDA and HBA. Besides, the
network is more fragile to non-adaptive MBA attack than to adaptive
HDA attack and this is due to its highly modular nature.

\begin{figure}
\begin{center}
\includegraphics[width=0.7\columnwidth]{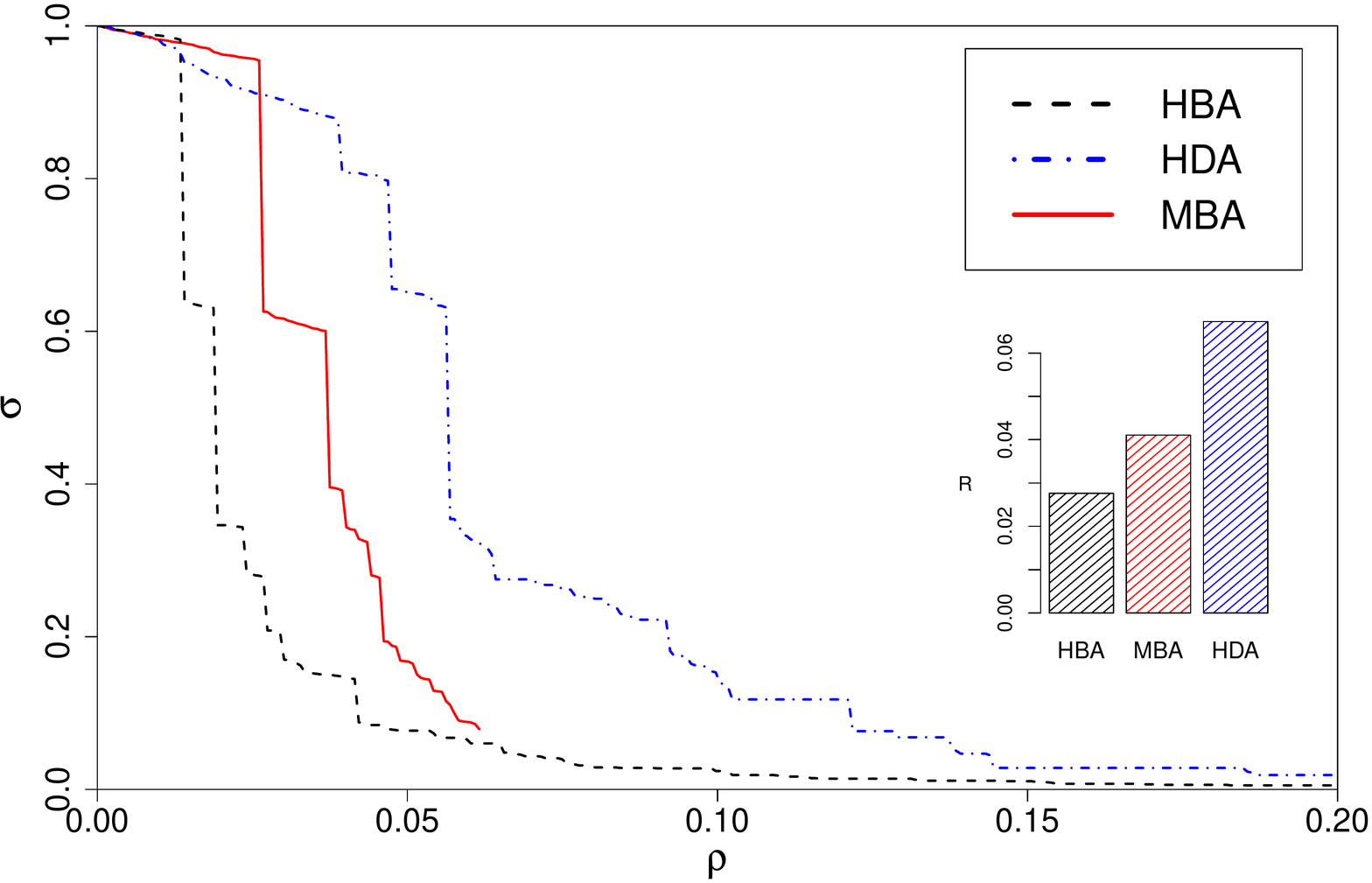}
\caption{The figure shows de effect of attacking the European Union
  power grid with HDA, HBA and MBA strategies measured by the relative size of the largest connected component
  $\sigma$ as a function of the fraction of removed nodes $\rho$. The inset displays the
  robustness of the network to all three attack strategies.}
\end{center}
\end{figure}\label{eupg}

\section{General discussion and conclusion}
In this contribution we have introduced a generalized robustness
measure and an empirical performance quantity to measure the trade-off
between computation time and robustness of modular networks facing
general attack strategies.  The two concepts were tested for a variety
of well known homogeneous and heterogeneous benchmark networks which
were attacked according to high degree adaptive attack, high
betweenness adaptive attack and non-adaptive module-based attack.
Computer simulations show that the module-based prescription perform
better than degree and betweenness sequential attacks for highly
modular networks ---for instance, the European power grid the
performance of the MBA is almost 10 times higher than the performance
of HDA and almost 6 times higher than the performance of HBA. This
outstanding result means that networks with well defined communities
present robustness to MBA and HBA very similar but with less
computational effort needed to atomize the network by the MBA
procedure.  Besides, the networks studied are more fragile to MBA than
to traditional sequential hub attack. These outstanding features
highlight the importance of interconnected nodes in maintaining
modular networks functioning as a whole.  Likewise, we have introduced
the concept of deactivation point where the network loses its
functionality and modular structure, which generally happens much
earlier than the percolation threshold usually used to quantify
network robustness.

Finally, we have studied the performance $\mathcal{P}$, the
deactivation point $P_{d}$, and the robustness for all three methods
described above for a real network example: the European power grid.
This network is highly modular and the attacks tried on it confirm
that the non-adaptive MBA procedure performs better than both HDA and
HBA.  Therefore, in order for the power grid system be safer against
malicious attacks, the modularity would decrease by, for instance,
rewiring internal edges in order to increase the number of
interconnected nodes.

The work is well posed as simulations were performed in a wide range
of benchmark networks with varying topologies at which the MBA
procedure is known to work very well.  We believe that these results
might have strong impacts in improving the robustness of real networks
and/or in planning effective attack strategies to real systems.

\section{Acknowledgments}
SG acknowledges financial support from brazilian agency Conselho
Nacional de Desenvolvimento Científico e Tecnológico through CNPq-MIT
project \#551974 /2011-7.  BRC acknowledges the brazilian Federal
Police for financial support.

\bibliography{bib_phd}

\end{document}